\documentclass[superscriptaddress,twocolumn,showpacs,preprintnumbers,amsmath,amssymb,prl,10pt,aps]{revtex4-1}
\usepackage{graphicx}
\usepackage{dcolumn}
\usepackage{bm}
\usepackage{subfigure}
\usepackage{multirow}
\usepackage{soul}
\usepackage{float}
\usepackage[normalem]{ulem}
\usepackage[usenames,dvipsnames]{color}
\usepackage{bbold}

\definecolor{tangerine}{rgb}{0.944,0.522,0}

\newcommand{\editor}[2]{%
  \expandafter\newcommand\csname #1note\endcsname[1]{%
    \textcolor{#2}{(\textbf{#1:} ##1)}}%
  \expandafter\newcommand\csname #1\endcsname[1]{%
    \textcolor{#2}{##1}}%
  \expandafter\newcommand\csname #1cancel\endcsname[1]{%
    \textcolor{#2}{\sout{##1}}}%
  \expandafter\newcommand\csname #1change\endcsname[2]{%
    \textcolor{#2}{\sout{##1} ##2}}%
  \newenvironment{#1text}{\color{#2}}{\color{black}}
}

\begin{document}

\title {Accurate modeling of FeSe with screened Fock exchange and Hund's metal correlations}
\author{Tommaso Gorni}
\affiliation{LPEM, ESPCI Paris, PSL Research University, CNRS, Sorbonne Universit\'e, 75005 Paris France}
\author{Pablo Villar Arribi}
\affiliation{LPEM, ESPCI Paris, PSL Research University, CNRS, Sorbonne Universit\'e, 75005 Paris France}
\author{Michele Casula}
\affiliation{Institut de Min\'eralogie, de Physique des Mat\'eriaux et de Cosmochimie (IMPMC), Sorbonne Universit\'e, CNRS UMR 7590, IRD UMR 206, MNHN, 4 Place Jussieu, 75252 Paris, France}
\author{Luca de' Medici}
\affiliation{LPEM, ESPCI Paris, PSL Research University, CNRS, Sorbonne Universit\'e, 75005 Paris France}

\date{\today}

%We reproduce the electronic properties of FeSe in the high-temperature phase within an \emph{ab initio} framework that includes screened Fock exchange and local dynamical correlations. We capture robustly the experimental band structure, as long as the system is in the Hund's metal phase. In particular we account for the shrinking of the Fermi pockets and the sinking below the Fermi level of the central pocket with atomic $xy$ orbital character. This entails the elusive correct estimate of the Sommerfeld coefficient, and supports the interpretation of non-compensated Fermi pockets seen in ARPES in terms of surface electron doping. More stringently, our modeling matches well the experimental interband optical spectrum, for which the correct description of each band's orbital character is paramount, and captures qualitatively the temperature dependence of the thermoelectric power, extremely sensitive to the details of the bands around the Fermi level.

\begin{abstract}
We reproduce the electronic properties of FeSe in the high-temperature phase within an \emph{ab initio} framework that includes screened Fock exchange and local dynamical correlations.
We robustly capture the experimental band structure, as long as the system is in the Hund's metal phase. 
In particular, we account for the shrinking of the Fermi pockets and the sinking  below the Fermi level of the hole pocket with $xy$ orbital character.
This entails the elusive correct estimate of the Sommerfeld coefficient, and supports the interpretation of non-compensated Fermi pockets seen in ARPES in terms of surface electron doping.
More stringently, our modeling matches well the experimental interband optical spectrum, 
%
%for which the correct description of each band's orbital character is paramount, 
%
and captures qualitatively the temperature dependence of the thermoelectric power, extremely sensitive to the details of the bands around the Fermi level.
\end{abstract}

\pacs{}

\maketitle

More than ten years after their discovery~\cite{Kamihara:2008}, consensus on the physics of iron-based superconductors (IBSC) is yet to be reached.
Albeit there are substantial indications pointing towards a spin-fluctuation pairing mechanism~\cite{Mazin_Splusminus,Hirschfeld_GapSymmetry,Chubukov_ItinerantScenario} for superconductivity induced by the proximity to magnetic instabilities, many important features are still not understood,
such as the nature of the nematic symmetry breaking, occurring in the normal state of many compounds, or the prediction of material trends.
In order to reach these goals, an accurate material-specific description of their electronic structure is essential.

Density-functional theory (DFT) in its standard implementations (the local-density approximation, LDA, or the generalized gradient approximation, GGA) qualitatively reproduces the structure of the Fermi surface, which is that of a semimetal (compensated in the parent compounds) with hole pockets in the center of the Brillouin zone and electron pockets at the corners%
%, for the parent compounds
~\cite{Zhang_Book_FeSC_ARPES}. However, when the calculated Fermi surfaces are compared with angle-resolved photoemission spectroscopy (ARPES) or Quantum Oscillation measurements, mismatches are found throughout all the IBSC families~\cite{Coldea:2008,Brouet:2009,Lee:2012,Terashima:2013,Watson:2015}.
Moreover, the DFT band structures are much more dispersing (at least a factor 2 to 3) than in experiments, and rescaling the bandwidth is not enough to bring the DFT band structure in agreement with the experimental findings~\cite{Ding_Arpes_BaK}.

This bandwidth mismatch can be ascribed to the lack of local dynamical correlations in DFT.
Indeed, local interactions and in particular Hund's coupling - the intra-atomic exchange - induce a strongly-correlated metallic phase in IBSC, which is dominated by fluctuating high-spin local configurations and therefore dubbed as a Hund's metal~\cite{Yin:2011,Georges_Annrev,demedici_Hunds_metals}. 
Electronic correlations in this phase strongly renormalize the bands in an orbitally-differentiated~\cite{demedici_3bandOSMT,Aichhorn:2010,demedici_MottHund,Yin:2011,YuSi_LDA-SlaveSpins_LaFeAsO,deMedici:2014}, thus highly non-trivial, way.
For this reason Dynamical mean-field theory (DMFT), and even the simpler Gutzwiller or slave-particle methods, drastically improve the overall band structure, and more generally electronic and magnetic properties~\cite{Hansmann_localmoment_prl,Yin:2011,Werner_122_dynU}.

LDA or GGA + local correlations, however, are not able to explain or reproduce some fundamental features, e.g.\ the size of electron and hole pockets of the Fermi surface~\cite{Yin:2011}. This issue is particularly striking in tetragonal FeSe, whose experimentally-measured hole pocket is from five to six times smaller than the one predicted by both standard DFT or DFT+DMFT~\cite{Watson:2015,Aichhorn:2010,Leonov:2015,Watson:2017}.
%
%Its normal phase does not display any long-range magnetic ordering when undergoing a tetragonal-to-orthorombic transition around~90\,K at ambient pressure~\cite{McQueen:2009}. The transition is believed to be driven by an electronic nematic instability, as pointed out by photoemission~\cite{Watson:2015}, or transport and NMR measurements~\cite{Baek:2015,Bohmer:2015}, hinting at a non-trivial behaviour of electronic correlations in the normal phase.
%
%In the context of weak-coupling model Hamiltonians, this deficiency has been related to the approximate treatment of non-local interactions in standard DFT.
%~\cite{Ortenzi:2009,Fanfarillo:2016,Jiang:2016,Scherer:2017},and some recent studies have shown that non-local self-energies improve the \emph{ab initio} prediction of the Fermi surface shape and low-energy excitations of LiFeAs~\cite{Zantout:2019,Bhattacharyya:2020}.
%
%A more accurate treatment of these is able instead to induce a relative shift in energy between the electron and hole pockets.
%s
Two main mechanisms have been proposed to cure this inaccuracy, one mediated by dynamical fluctuations~\cite{Ortenzi:2009,Fanfarillo:2016}, the other as a static effect of intersite Coulomb repulsion~\cite{Jiang:2016,Scherer:2017} treated at the Hartree-Fock level.
Up to now, the lack of a detailed comparison with experiments of an \emph{ab initio} description of FeSe's normal phase has not allowed to clearly discriminate between the two scenarios.

In this Letter, we tackle the \emph{ab initio} modeling of FeSe and we show that an accurate description of the Fermi surfaces can be obtained by including screened Fock exchange effects in the reference Hamiltonian via \emph{hybrid functional}-DFT~\cite{Marsman:2008,Franchini:2014}.
Moreover by including as well local dynamical correlations due to the Hund's metallicity we obtain a quasiparticle band structure mathching quite accurately (both in dispersion and orbital character) ARPES and transport experimental measurements. 
Our result thus supports the static mechanism for the shrinking of the Fermi pockets in this material, and more in general the non-trivial disentanglement of the electronic self-energy in a static and non-local part, and a dynamical local one:
\begin{equation}
\Sigma({\bf k},\omega)
=
\Sigma_{\rm non-loc}({\bf k})+\Sigma_{\rm dyn}(\omega),
\end{equation}
to a good level of approximation, in line with earlier GW calculations on IBSC~\cite{Tomczak:2012,Tomczak:2015} and oxides~\cite{Tomczak:2014}, and a more recent analysis of ARPES experimental data~\cite{Kim:2020}.

\paragraph*{Methods.}
%\label{sec:methods}

We calculate a reference one-body DFT Hamiltonian with the HSE hybrid functional~\cite{Heyd:2003,Heyd:2006}, which includes 
%a static screened-exchange type of interaction among the Kohn-Sham electrons.
%, and moderates the LDA and GGA tendency to over-delocalize the electrons~\cite{Marsman:2008,Franchini:2014}. 
%In practice, this is accomplished by including 
a fraction of the short-range Hartree-Fock exchange energy into the standard Perdew-Burke-Ernzerhof (PBE)~\cite{Perdew:1996} GGA functional.
Throughout this work we compare our results with those obtained using the standard PBE functional.
All the DFT calculations have been carried out using the experimental lattice parameters $a = 3.7707 \AA$, $c=5.521 \AA$ and $z_{\rm Se} = 0.2667 $ for the tetragonal phase of FeSe~\cite{Margadonna:2008,Lehman:2010}, using the {\sc Quantum ESPRESSO} package~\cite{Giannozzi:2017}~(see Supplementary Material for the numerical details).
%
%The non-local exchange operator present in HSE has been treated with the ACE scheme to speed-up the calculation~\cite{Carnimeo:2019,Lin:2016} 

As in all IBSC the bands crossing the Fermi level in FeSe are mainly of Fe-$3d$ character, and in order to include  many-body effects through the explicit treatment of local correlations a parametrization of the one-body Hamiltonian on a localized basis is necessary.
We thus project the DFT Kohn-Sham Hamiltonian for these bands over a set of five maximally-localized Wannier functions per site, using the {\sc wannier90} code~\cite{Pizzi:2020}. This downfolding process yields the hopping and on-site energies of the one-body part
\begin{equation}
\hat{\mathcal{H}}_0
=
\sum_{\substack{i\neq j \\ m,m',\sigma}}
t^{mm'}_{ij}
\hat{d}^{\dag}_{im\sigma}
\hat{d}_{jm'\sigma}
+
\sum_{i,m,\sigma}
\epsilon_{im\sigma}
\hat{n}_{im\sigma}
\, ,
\label{eq:H0}
\end{equation}
where $\hat{d}^{\dag}_{im\sigma}$ creates an electron in the Wannier spin-orbital $m\sigma$ and lattice site $i$, and $\hat{n}_{im\sigma} = \hat{d}^{\dag}_{im\sigma}\hat{d}_{im\sigma}$ is the corresponding number operator. 

A Hubbard-Kanamori Hamiltonian is used to include the local interactions to be treated in a dynamical many-body fashion
\begin{align}
\hat{\mathcal{H}}%-\mu\hat{N} 
=\: &
\hat{\mathcal{H}}_0 %-\mu \sum_{m\sigma} \hat{n}_{m\sigma}
+
U
\sum_{im} \hat{n}_{im\uparrow}\hat{n}_{im\downarrow}
\nonumber\\
&
+
(U-2J)
\sum_{i,m\neq m'} \hat{n}_{im\uparrow}\hat{n}_{im'\downarrow}
\nonumber\\
&
+
(U-3J)
\sum_{i,m < m',\sigma} \hat{n}_{im\sigma}\hat{n}_{im'\sigma}
\, ,
\label{eq:kanamori}
\end{align}
where U is the intra-orbital Coulomb repulsion and J the Hund's coupling~\footnote{Here the local off-diagonal terms of the Kanamori Hamiltonian (spin-flip and pair-hopping) are customarily dropped.}. In this framework the double counting energy for these interactions is absorbed in the chemical potential.

Here we solve the many-body Hamiltonian~\eqref{eq:kanamori} within the slave-spin mean-field theory (SSMF)~\cite{demedici_Slave-spins,demedici_Vietri}, which is simpler, but also a lot cheaper than DMFT, yet provides a $\Sigma_{\rm dyn}(\omega)$ which has proven to be a robust approximation for IBSC by, e.g., successfully capturing their orbital-differentiation signatures~\cite{deMedici:2014}, or predicting the evolution of the Sommerfeld coefficient upon doping  in the 122 family~\cite{Hardy:2016}.
SSFMT describes the Fermi-liquid low-temperature paramagnetic metallic phase of~\eqref{eq:kanamori} as a quasiparticle Hamiltonian
\begin{equation}
\hat{\mathcal{H}}_{\rm QP}
=
\!\!\!
\sum_{\substack{i\neq j\\ mm'\sigma}}
\!\!
\sqrt{Z_m Z_{m'}}
\,
t^{mm'}_{ij}
\hat{f}_{im\sigma}^{\dag}
\hat{f}_{jm'\sigma}
+
\sum_{im\sigma}
(\epsilon_m -\tilde{\lambda}_m)\hat{n}^f_{im\sigma}
\, ,
\label{eq:H-QP}
\end{equation}
where the (orbital-dependent) quasiparticle renormalizations $Z_m$ and on-site-energy shifts $\tilde{\lambda}_m$ are determined solving the self-consistent slave-spin equations for given values of the local interactions $U$ and $J$ (see~\cite{demedici_Vietri} and Supplementary Material).

\begin{figure}
\centering
\includegraphics[width=\columnwidth,trim=0 0 20 0 clip]{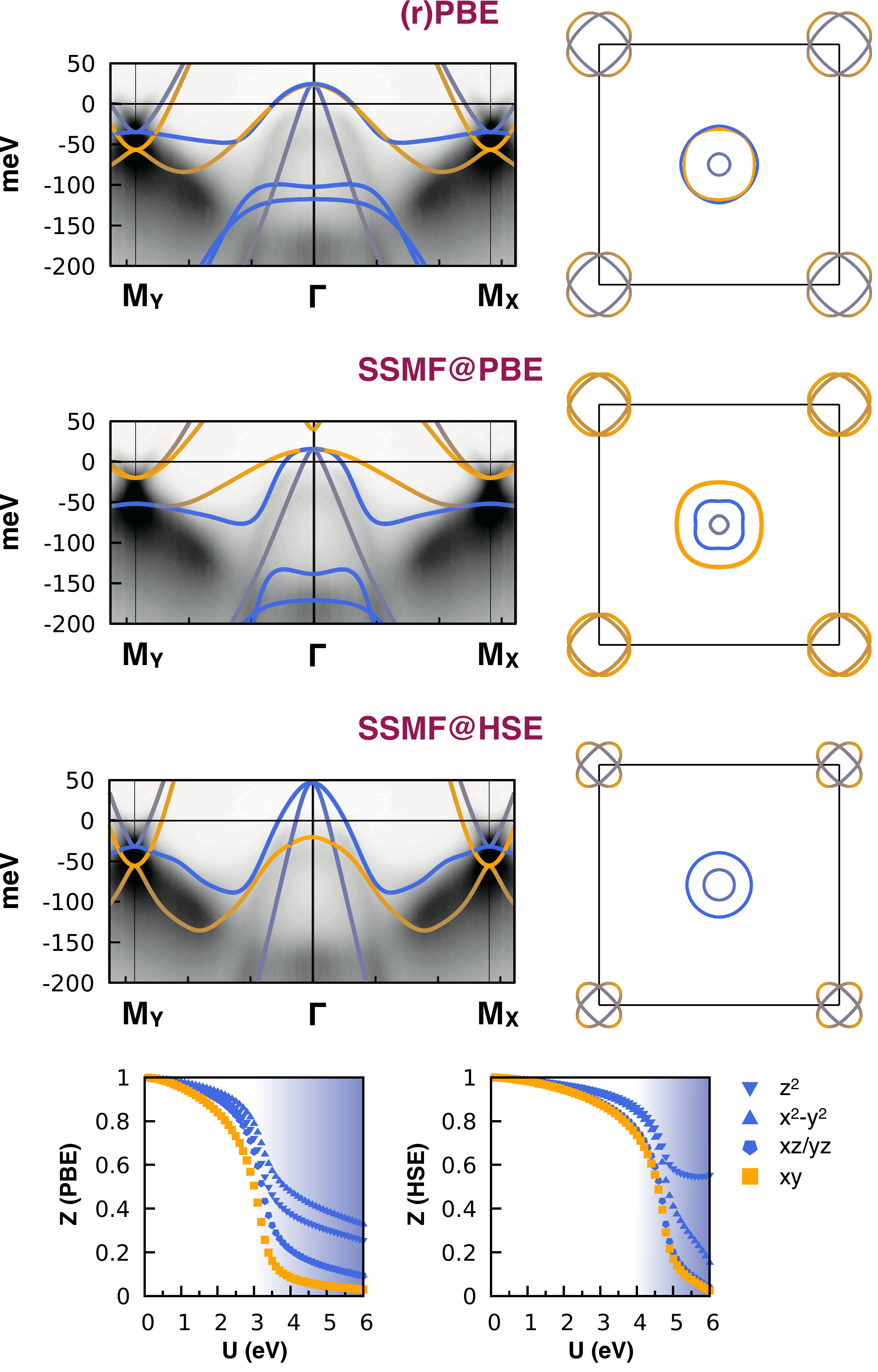}
\caption{%
Calculated band dispersion along the $M_Y$-$\Gamma$-$M_X$ path (solid lines) on top of the corresponding ARPES dispersions from Ref.~\cite{Huh:2020} (grey scale),
and $k_z\!\!=\! 0$-Fermi surface of FeSe, shown in the 2-Fe Brillouin Zone. The $xy$ orbital weight is represented in a colour scale going from blue (zero weight) to orange (maximum weight).
Upper panel: PBE case renormalized by a constant factor $1/Z=6$. 
Upper-middle panel: SSMF@PBE solution with local interaction parameters $U=4.2$\,eV and $J/U=0.20$.
Lower-middle panel: SSMF@HSE solution with local interaction parameters $U=5.0$\,eV and $J/U=0.20$. Only the latter reproduces correctly the shrinking of the Fermi pockets, the sinking of the $xy$-band at $\Gamma$ and the band dispersions.
In the two bottom panels, the SSMF mass renormalizations for fixed $J/U=0.20$ are reported for both the reference Hamiltonians considered. The Hund's metal region is shaded in blue.
\label{fig:bands}
}
\end{figure}

When using the PBE reference $H_0$ (we label the final result SSMF@PBE), we set $U^{\rm PBE} = 4.2$\,eV and $J^{\rm PBE} = 0.2\,U^{\rm PBE}$, based on constrained random-phase approximation calculations~(cRPA~\cite{Miyake:2010}) and benchmarking on higher-level theories (see Supplemental Material). In the HSE case (labeled SSMF@HSE), we keep the same $J/U$-ratio as for PBE, and we fix $U^{\rm HSE} = 5.0$\,eV, the value at which the $t_{2g}$ mass renormalization is closest to the SSMF@PBE one.
This choice is further validated by performing several scans in the $(U,J)$-space, showing that the main improvements with respect to the SSMF@PBE case are robust with respect to changes in the local interactions, as long as the system is within the Hund's-metal phase. 
This is indeed a region of strong and orbitally-differentiated mass renormalization, which is systematically found in these models\cite{demedici_Hunds_metals,Lanata_FeSe_LDA+Gutz,VillarArribi_FeSe_el_comp,Liebsch_FeSe_spinfreezing,Werner_SpinFreezing,Fanfarillo_Hund} at interaction strength beyond a cross-over value (see blue-shaded region in bottom panels in Fig. \ref{fig:bands}).

%==========================================================
%
\paragraph*{Quasiparticle dispersion and Fermi surface.}
%\label{sec:bands}
%
%==========================================================
%
The itinerant-fermion description of FeSe yielded by the PBE (or LDA) functional is known to predict an overall correct shape of the Fermi surface, but also to overestimate the hole-pocket size of about a factor 5 and underestimate the effective masses up to 8 times~\cite{Watson:2015,Coldea:2018}. 
For this reason, the PBE bands shown in the upper panel of Fig.~\ref{fig:bands} are renormalized by a constant factor $Z=1/6$, so to ease the comparison of their dispersion with the subsequent cases. It must be noted, however, that this simple rigid renormalization does not account for the quasiparticle mass differentiation reported in experiments, with values of $m^*/m_b=1/Z$ ranging from $2$--$3$ for the $xz/yz$ bands up to $8$ for the $xy$ bands, with $m_b$ being the PBE masses~\cite{Watson:2015}.
The mass differentiation is naturally brought in by the dynamical local correlations in the Hund's metal region along with the strong overall renormalization, as shown by %DMFT~\cite{Aichhorn:2010,Liebsch_FeSe_spinfreezing}, Gutzwiller~\cite{Lanata_FeSe_LDA+Gutz} and, equivalently, by 
our SSMF@PBE calculations presented in the upper-middle panel in Fig.~\ref{fig:bands}. 
However, regardless of the methodology used, local self-energies alone do not improve the PBE Fermi surface, yielding Fermi pockets of roughly the same size as PBE and an overrepresentation of the $xy$ character~\cite{Aichhorn:2010,Leonov:2015,Watson:2017}, in stark contrast with experimental evidence~\cite{Watson:2015,Yi:2019,Huh:2020}.

The main outcome of this work is reported in the lower-middle panel of Fig.~\ref{fig:bands}, displaying the band structure and Fermi surface obtained with SSMF when using the HSE reference Hamiltonian.
When the HSE static screened Fock exchange is included in the quasiparticle equations, the $xy$ band sinks below the Fermi level at the zone centre, leaving a smaller hole pocket of $xz/yz$ character and explaining the origin of the flat band seen in ARPES  about 50\,meV below $\varepsilon_{\rm F}$~\cite{Watson:2015,Suzuki:2015}. 
At the zone corner, a relevant $xy$ weight remains in the outermost part of elliptical pockets, in agreement with recent polarized-ARPES data~\cite{Huh:2020}, and the $xy$ and $xz/yz$ bands form together a double-hourglass shape with bands bottoms at $60$ and $30$\,meV below $\varepsilon_{\rm F}$, consistently with their experimental position of about $50$--$60$\,meV and $20$--$40$\,meV below $\varepsilon_{\rm F}$, respectively~\cite{Nakayama:2014,Watson:2015,Zhang:2015,Fanfarillo:2016,Yi:2019}.

A crucial consequence of this band rearrangement is the net reduction of the size of the Fermi pockets, pointing towards static non-local Coulomb effects as the most likely pocket-shrinking mechanism in FeSe. Furthermore, the competing dynamical non-local scenario has been recently shown to be quite ineffective in FeSe, even when accounting for realistic non-local dynamical fluctuations~\cite{Bhattacharyya:2020}.
%
%The overestimation of the Fermi pocket size of IBSC in \emph{ab initio} descriptions, most severe in FeSe, has been related to the lack of non-local interactions previously~\cite{Ortenzi:2009,Jiang:2016}, arguing they can induce the needed interband repulsion between the electron and hole pockets.
%
%Two main mechanisms have been proposed in the context of weak-coupling model Hamiltonians, one mediated by dynamical fluctuations~\cite{Ortenzi:2009,Fanfarillo:2016}, the other as a static effect of intersite Coulomb repulsion~\cite{Jiang:2016,Scherer:2017}. 
%
%Up to now, the lack of a proper \emph{ab initio} description of FeSe normal phase has not allowed to clearly discriminate between the two scenarios.
%
%The success of our approach in the description of the pocket shrinking clearly points towards static non-local interactions as the main responsible for such an effect in FeSe. Moreover, a recent study has shown the dynamical non-local scenario to be quite ineffective in FeSe, even when accounting for realistic non-local dynamical fluctuations~\cite{Bhattacharyya:2020}.
%
We remark that the observed Fermi surface shrinking is a multiband effect, with the $xz/yz$ pockets being more sensitive to local dynamical correlations and the $xy$ pockets more affected by non-local static effects, as long as the local correlations are strong enough to place FeSe in the Hund's-metal phase (see Supplemental Material). 

%
% Compensation
%
Finally, a relevant point that has never been explicitly addressed in the literature to our knowledge is the lack of compensation (between the volumes of the hole and electron pockets, due to the Luttinger theorem for Fermi liquids) inferred from ARPES measurements (see Tab.~1 in the Supplemental Material).
The experimentally reported volumes of the $xz/yz$ electron and hole pockets tend to compensate one another, therefore the volume of the $xy$ electron pocket is expected to be compensated by a corresponding hole pocket, which however has always escaped detection.
The general difficulty in resolving the $xy$ bands in ARPES~\cite{Fanfarillo:2016}, together with the prediction of a high $xy$ scattering rate~\cite{Aichhorn:2010}, may lead to the conjecture that a faint, undetected $xy$ hole pocket is indeed present at the zone centre.
However, the stark improvement in our calculated band structure produced by the $xy$ hole pocket sinking below the Fermi level, together with the clear experimental detection of an $xy$ band in the same energy range in the low temperature phase~\cite{Watson:2015}, lead us to discard the occurrence of an $xy$ hole pocket in FeSe, and to propose surface electron doping as the most likely explanation to the lack of compensation in ARPES experiments.
A precise quantification of this doping would necessitate to account for finer energy scale effects such as spin-orbit coupling and is beyond the scope of this study. We thus limit ourselves to estimate that a rigid chemical potential shift of the SSMF@HSE band structure reproduces the experimental pocket sizes for an electron doping of about 0.1$e^-$ per Fe.

%==========================================================
%
\paragraph*{Specific heat.}
%\label{sec:Sommerfeld}
%
%==========================================================
%
%
% Sommerfeld (better put before the estimate or the theoretical results?)
%
We now turn to the Sommerfeld coefficient $\gamma_n$, i.e.\ the linear coefficient in the electronic specific heat as a function of temperature. The Sommerfeld coefficient is proportional to the bulk density of states (DoS) at the Fermi level, which is ultimately a sensitive probe of the renormalized band structure and thus of electronic correlations~\cite{Ashcroft}.
Even though the FeSe tetragonal phase is not stable below $90$\,K, hampering a precise extrapolation of the specific heat to zero temperature, a rough estimate of its Sommerfeld coefficient can be inferred based on the low-temperature phase measurements, leading to a value of about $\gamma_n \approx 12$\,mJ\,mol$^{-1}$K$^{-2}$~\cite{Hardy:2019}\footnote{F. Hardy, private communication.}.
Comparing with the ones predicted theoretically within the different schemes, SSMF@HSE yields $\gamma_n=9.0$\,mJ\,mol$^{-1}$K$^{-2}$, in much closer agreement with the experimental estimate than the SSMF@PBE value of $\gamma_n=36.6$\,mJ\,mol$^{-1}$K$^{-2}$, or than the (unrenormalized) PBE value of  $\gamma_n=4.0$\,mJ\,mol$^{-1}$K$^{-2}$.
The difference between the two correlated cases can again be traced back to the $xy$ band, whose positioning below the Fermi level is further supported by this estimation.

% SOC ?
%
% wrong topology of xy band at M?

\paragraph*{Transport properties.}
%\label{sec:optic}

\begin{figure}[h] 
\centering
\includegraphics[width=\columnwidth,trim=0 0 0 0 clip]{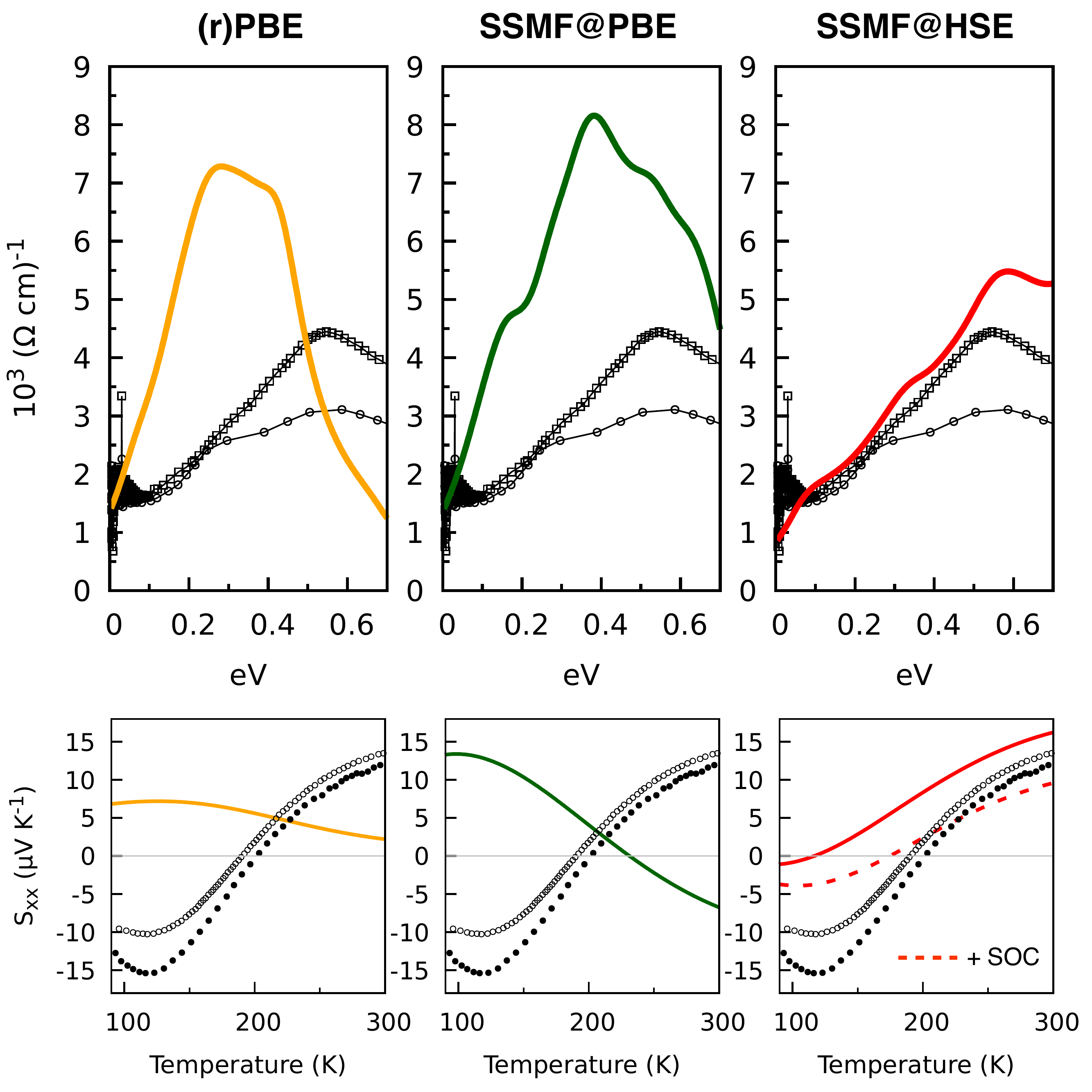}
\caption{In-plane interband contribution to optical conductivity (upper panels) and thermopower (lower panels) of FeSe, computed in the renomalized-PBE, SSMF@PBE and SSMF@HSE cases.  
Optical conductivity calculations (solid lines, upper panels) are compared with the experimental data of 205\,nm-FeSe@CaF$_2$ at $T = 100$\,K~\cite{Nakajima:2017} (open circles), and FeSe monocrystal al $T = 120$\,K~\cite{Wang:2016} (open squares), both subtracted of their coherent Drude peak. A Lorentzian broadening $\eta = 0.06$\,eV has been used in the computation of the theoretical spectra.
Seebeck calculations (solid lines, lower panels) are compared with experimental data of FeSe polycrystals (open circles from~\cite{Lodhi:2019}, full circles from~\cite{Song:2011}), for the temperature range above the nematic transition, occurring around $T= 90$\,K at ambient pressure~\cite{McQueen:2009}.
\label{fig:opt-cond}
}
\end{figure}  

In order to assess the quality of our band structure over a broader energy range, we compute the optical conductivity of tetragonal FeSe. Neglecting vertex corrections, its real part reads as
\begin{align}
{\rm Re}\Big[\sigma_{\alpha\beta}(\omega)\Big]
&=
\frac{2\pi e^2 \hbar}{\mathcal{V}}
\sum_{\mathbf{k}}^{\rm BZ}
\int_{-\infty}^{+\infty}
{\rm d}\omega'\,
\frac{f(\omega') -f(\omega+\omega')}{\omega}
\nonumber\\
&\hspace{2cm}
\times
{\rm Tr}
\Big[
v_{\mathbf{k}\alpha}A_{\mathbf{k}}(\omega')v_{\mathbf{k}\beta}A_{\mathbf{k}}(\omega+\omega')
\Big]
\, ,
\label{eq:opt-cond}
\end{align}
where $f(\omega)$ is the Fermi function, $A_{\mathbf{k}}(\omega)$ is the spectral function matrix and $v^{mm'}_{\mathbf{k}\alpha}$ the velocity matrix elements computed within the Peierls approximation~\cite{Tomczak:2009,Valenzuela:2013}.
%
%Within the SSMFT approximation the self-energy is a purely real function, we then 
Within SSMF quasiparticles have infinite lifetime, which implies an unbroadened Drude peak in optics, an artificial feature in this context. We thus
examine only the interband contributions to the optical conductivity~\cite{Calderon-Optics_Pnictides}, so to avoid empirical fittings of the Drude width. In the upper panels of Fig.~\ref{fig:opt-cond} we report the interband contribution to the in-plane optical conductivity (i.e.\ in the Fe-Fe plane, along the Fe-Se direction) in the renormalized-PBE, SSMF@PBE, and SSMF@HSE case. 
Optical conductivity data in IBSC are usually interpreted by means of two different Drude peaks, describing respectively the coherent and  incoherent electron dynamics. We choose to subtract only the coherent Drude peak to the experimental data, since the experimentally-fitted incoherent one, with a broadening of 
% or hbar?
$h/\tau\approx 0.2$\,eV~\cite{Nakajima:2017}, might include portions of interband transitions.
From the upper panels of Fig.~\ref{fig:opt-cond} one can see that SSMF@HSE better reproduces the experimental trends, predicting a main absorption around $0.6$\,eV and a secondary one around $0.3$\,eV.
The higher-energy transition is detected both in the FeSe monocrystal~\cite{Wang:2016}, around $\hbar\omega_0 \approx 0.55$\,eV at $T = 120$\,K, and in  205\,nm-FeSe@CaF$_2$~\cite{Nakajima:2017},  around $\hbar\omega_0 \approx 0.62$\,eV at $T = 100$\,K, whereas a less intense interband transition is reported in 205\,nm-FeSe@CaF$_2$  around $\hbar\omega_0 \approx 0.25$\,eV at $T = 100$\,K.
On contrary, both the renormalized-PBE and SSMF@PBE optics yield an excessively intense absorption peak between $0.3$--$0.4$\,eV, highlighting the crucial role played by screened Fock exchange in reproducing not only the band structure in detail but also the bands orbital character, which shapes - through the velocity matrix elements - the interband optical conductivity.  
%pointing towards the importance of non-local interactions even for the hole-like excitations.
%
We finally notice that in the SSMF@HSE case a fainter interband absorption around $0.1$\,eV is present and not reported by either experiments. Its intensity and position suggest that it might have been incorporated into the empirically-fitted incoherent Drude peak, with a consequent overestimation of its intensity.

%
% Seebeck
%
The Seebeck effect, i.e.\ the induction of a charge current due to a thermal gradient, is a phenomenon sensitive to the details of the electronic structure within the thermal energy windows, making thermopower measurements a standard probe for Fermi surface reconstructions and related low-energy properties~\cite{Pallecchi:2016}.
For the same reason, it provides a very challenging playground for theoretical predictions, where even the qualitative agreement has not to be taken for granted~\cite{Mravlje_Thermopower}.
In a system showing no transverse current response, the thermopower is a diagonal tensor reading as 
\begin{equation}
S_{\alpha\alpha}(T)
=
-
\frac{1}{eT}
\frac
{\displaystyle \int {\rm d} \epsilon \big(\epsilon- \mu\big) \left(-\frac{\partial f}{\partial \epsilon}\right) \Xi_{\alpha\alpha}(\epsilon) }
{\displaystyle \int {\rm d}\epsilon \left(-\frac{\partial f}{\partial \epsilon}\right) \Xi_{\alpha\alpha}(\epsilon)}
\, ,
\label{eq:seebeck}
\end{equation}
where $e$ is the absolute value of the electron charge, and both the chemical potential $\mu$ and the derivative of the Fermi function $f(\epsilon)$ are meant at their temperature-$T$ value. The integrals in Eq.~\eqref{eq:seebeck} involve the transport distribution function $\Xi_{\alpha\beta}(\epsilon) ={\rm Tr}
\Big[
v_{\mathbf{k}\alpha}A_{\mathbf{k}}(\epsilon)v_{\mathbf{k}\beta}A_{\mathbf{k}}(\epsilon)
\Big]$,
that we evaluate within SSMF. 
This is a quite rough approximation, since it implies a constant relaxation time that simplifies between numerator and denominator in eq. (\ref{eq:seebeck}). Also we neglect phonon-drag effects.
The in-plane thermopower as a function of temperature is shown in the lower panels of Fig.~\ref{fig:opt-cond}, compared with experimental data on polycrystalline samples. 
%
% Discuss the sign?
%
Local correlations alone are unable to predict the correct trend and sign over the whole temperature range, whereas the inclusion of screened Fock exchange (SSMF@HSE) restores the correct behavior below room temperature, where the agreement with experiments is closer, and predicts a change of sign within the tetragonal phase when lowering the temperature.
The agreement with the experimental data deteriorates as $T$ decreases, which can be partially explained by the lack of spin-orbit coupling (SOC) in our calculations. Indeed, band splittings up to $20$\,meV have been related to SOC in FeSe~\cite{Watson:2015,Suzuki:2015,Borisenko:2016,Zhang:2016,Fanfarillo:2016}, enough to affect the low-temperature behaviour of the Seebeck coefficient (see dashed line in the last panel of Fig.~\ref{fig:opt-cond} and Supplementary Material).

\paragraph*{Conclusions.}
%\label{sec:conclusions}

In summary, we have shown that the inclusion of screened Fock exchange in DFT, through the hybrid functional HSE, and of local Hund's-metal correlations within the slave-spin mean field, yields a remarkably accurate description of the band structure of FeSe. Compared to the standard PBE approximation to DFT, using the HSE reference Hamiltonian induces a shrinking of the Fermi pockets, bringing their size closer to the experimental values, and generally improves the description of the renormalized quasiparticle bands and of transport properties.
We remark that our SSMF@HSE approach is arguably the numerically cheapest way to incorporate all the main physical effects necessary for an accurate description of FeSe, with respect to the more complete but computationally heavier GW+DMFT\cite{Tomczak:2012} or SEX+DMFT\cite{vanRoekeghem_SEX+DMFT}. 

Our results suggest that the lack of compensation in ARPES data may be due to surface electron doping, and point towards static short-range Coulomb effects to be the most likely mechanism for pocket shrinking in FeSe. 
Finally, our findings support the conjecture that non-local and dynamical effects can be disentangled in the self-energy of IBSC to a good level of approximation~\cite{Tomczak:2012,Tomczak:2015,Kim:2020}.

\section*{ACKNOWLEDGMENTS}
The authors acknowledge fruitful discussions with V.\ Brouet, L.\ Fanfarillo, A.\ Georges, F.\ Hardy, L.\ Paulatto, S.\ Pons.
TG, PVA and LdM are supported by the European Commission through the ERC-StG2016, StrongCoPhy4Energy, GA No724177. 
This work was performed using HPC resources from GENCI-IDRIS/TGCC (Grant A0060910777 and 0906493). 
This work was granted access to the HPC resources of MesoPSL financed by the Region \^Ile-de-France and the project Equip@Meso (reference ANR-10-EQPX-29-01) of the programme Investissements d'Avenir supervised by the Agence Nationale pour la Recherche.

\bibliography{FeSe_hybrid.bib,FeSc.bib,publdm.bib,bibldm.bib}

\end{document}

\title{Supplementary Material}
\author{Tommaso Gorni}
\affiliation{LPEM, ESPCI Paris, PSL Research University, CNRS, Sorbonne Universit\'e, 75005 Paris France}
\author{Pablo Villar Arribi}
\affiliation{LPEM, ESPCI Paris, PSL Research University, CNRS, Sorbonne Universit\'e, 75005 Paris France}
\author{Michele Casula}
\affiliation{Institut de Min\'eralogie, de Physique des Mat\'eriaux et de Cosmochimie (IMPMC), Sorbonne Universit\'e, CNRS UMR 7590, IRD UMR 206, MNHN, 4 Place Jussieu, 75252 Paris, France}
\author{Luca de' Medici}
\affiliation{LPEM, ESPCI Paris, PSL Research University, CNRS, Sorbonne Universit\'e, 75005 Paris France}

\maketitle

%===============================
%
\section{Simulation details and choice of the interaction parameters}
%
%===============================

\begin{figure}
\includegraphics[width=0.7\textwidth]{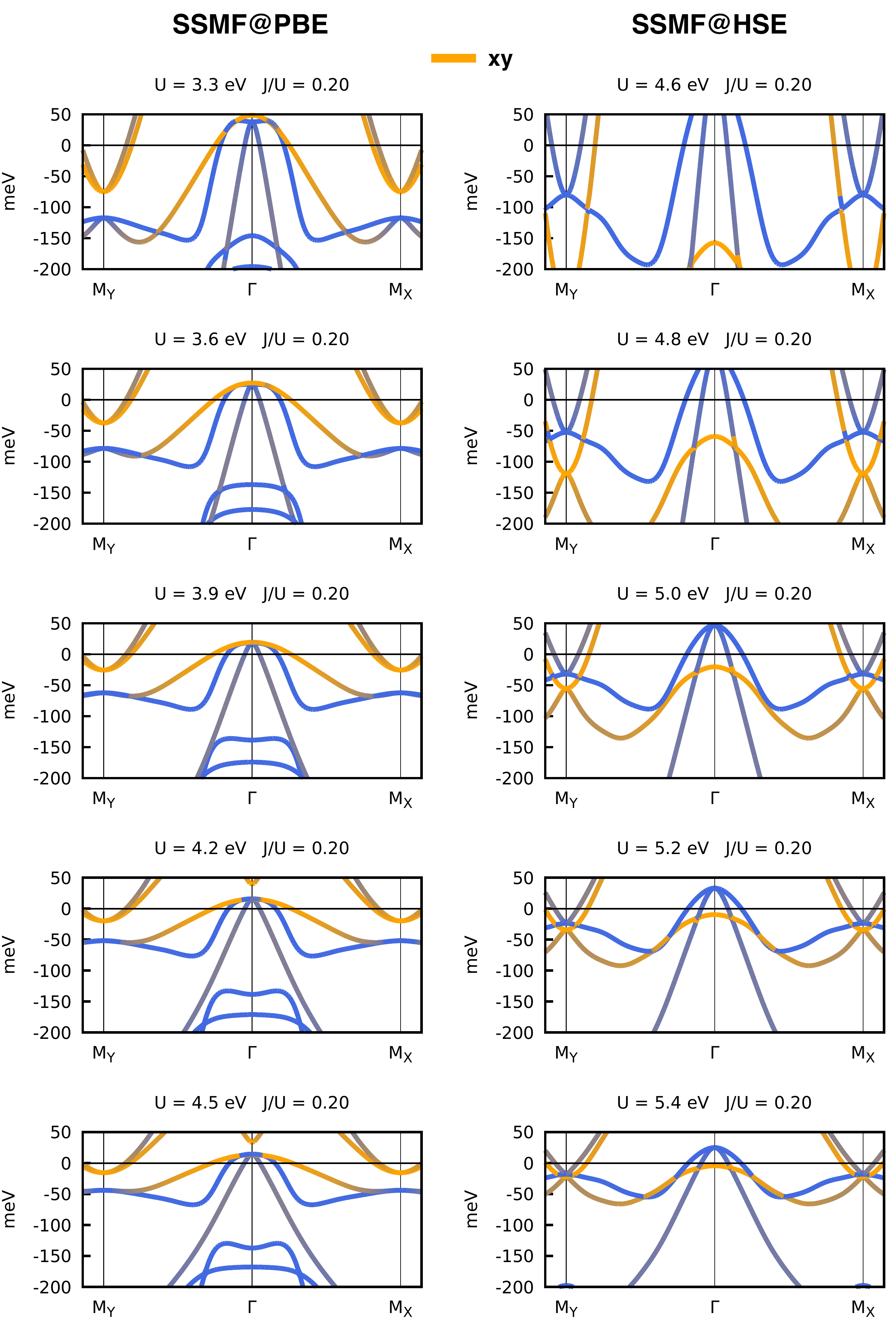}
\caption{Dependence of the position of the $xy$ band on the local interaction parameters for both the reference Hamiltonians considered.
\label{fig:bands-UJ}}
\end{figure}
 
\begin{figure}
\includegraphics[width=\textwidth]{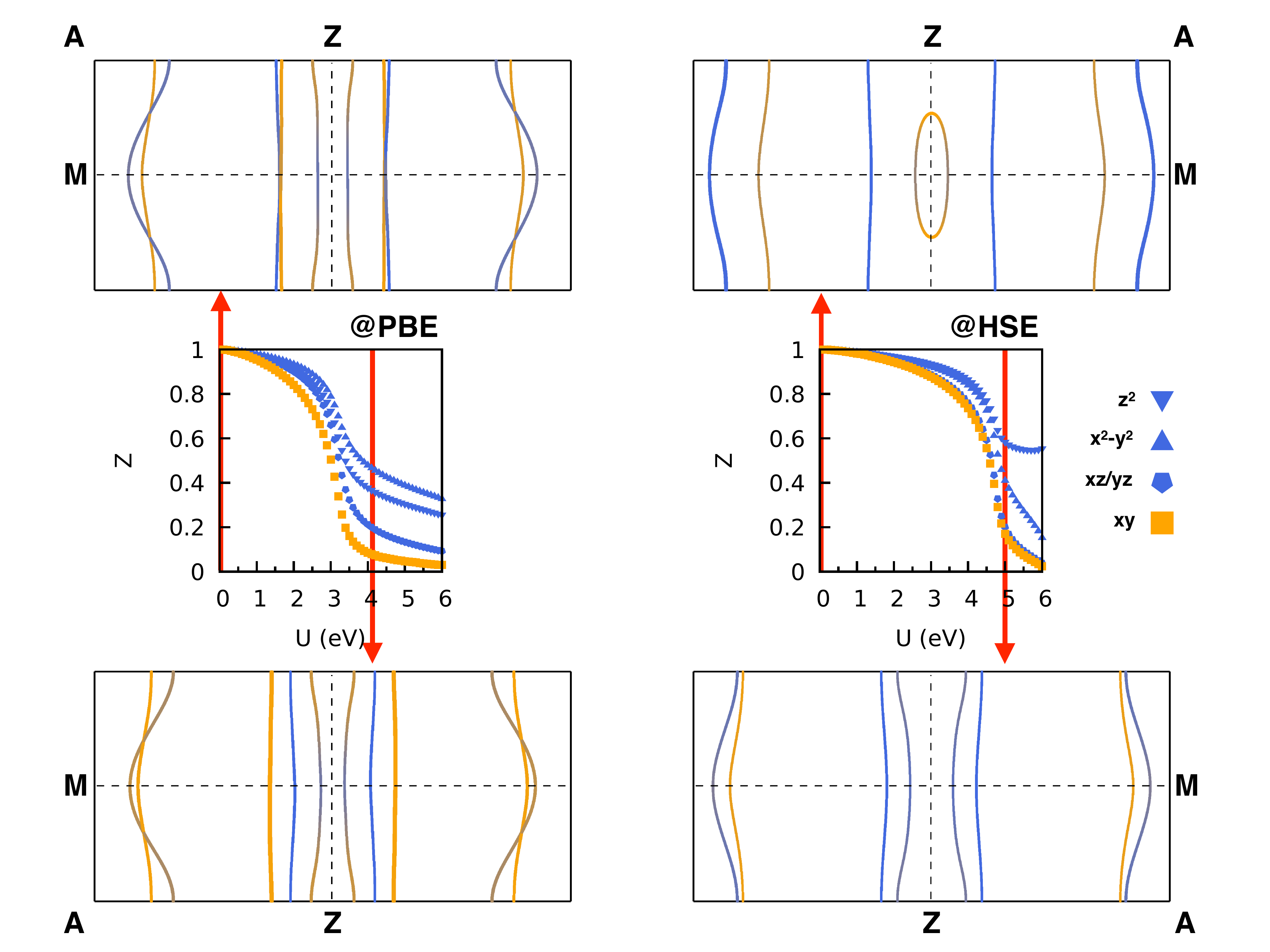}
\caption{Orbitally-resolved Fermi surface along the plane passing through the $\Gamma$,$Z$ and $M$ symmetry points in the four cases considered in this work, as indicated by the $Z_{\nu}(U)$ values in the inset ($Z_{\nu}=1$ corresponds to the pure-DFT case, whether it being PBE or HSE).
\label{fig:FS-GZM}}
\end{figure}

%
% DFT
%
The DFT calculations with the PBE and HSE functional have been carried out with the {\sc Quantum ESPRESSO} package~\cite{Giannozzi:2009,Giannozzi:2017}, using the PBE norm-conserving pseudopotentials from the PseudoDojo library~\cite{vanSetten:2018} and a plane-wave cutoff of $\epsilon_{\rm cut} = 90$\,Ry. Integrals over the Brillouin zone have been converged with a gaussian smearing of $\sigma = 0.01$\,Ry and a Monkhorst-Pack grid of $8\times 8 \times 8$ points. 
The Fock exchange operator present in HSE has been treated within the adaptively compressed exchange (ACE) scheme to speed-up the calculation~\cite{Carnimeo:2019,Lin:2016}. It enters the Kohn-Sham equations in the form of
\begin{equation}
v_{\rm sex} (\mathbf{r},\mathbf{r}')
=
- \gamma(\mathbf{r},\mathbf{r}') 
W^{\rm HSE}(|\mathbf{r}-\mathbf{r}'|)
\, ,
\label{eq:sex}
\end{equation}
where $\gamma(\mathbf{r},\mathbf{r}')$ is the one-particle-reduced density matrix $ \gamma(\mathbf{r},\mathbf{r}') =\sum_{n\mathbf{k}}f_{n\mathbf{k}}\psi_{n\mathbf{k}}(\mathbf{r})\psi^*_{n\mathbf{k}}(\mathbf{r}')$, with $\psi_{n\mathbf{k}}$ being the Kohn-Sham states and $f_{n\mathbf{k}}$ their occupations, and $W^{\rm HSE}(|\mathbf{r}-\mathbf{r}'|)$ a screened Coulomb potential in the form of~\cite{Franchini:2014,Aryasetiawan:1998}
\begin{equation}
W^{\rm HSE}(r)
=
e^2 \frac{\alpha\, {\rm erfc}(k_c r)}{r}
\, .
\label{eq:hse-screening}
\end{equation}
In the latter equation,  ${\rm erfc}(x) = 1 - {\rm erf}(x)$ is the complementary error function, $\alpha=0.25$ the screened-exchange fraction included in the DFT energy functional and $k_c=0.2\,\AA^{-1}$ the screening parameter, corresponding to the critical distance $r_s \approx 10\AA$, at which the semilocal PBE~\cite{Perdew:1996} exchange takes over.

%
%  Wannier
%
The band structure for the HSE functional has been obtained by interpolating the Kohn-Sham eigenvalues by means of maximally-localized Wannier functions spanning the Fe-$3d$ and Se-$4p$ manifold ($pd$ model). The tight binding model used in SSMF calculations has been obtained with Wannier functions spanning only the Fe-$3d$ manifold, consistently with what done in the PBE case ($d$ model). The interpolation of the $d$-manifold yielded by the $d$ model has been found to be identical to the one yielded by the $pd$ model both in the PBE and the HSE case.

\begin{figure}
\includegraphics[width=\textwidth]{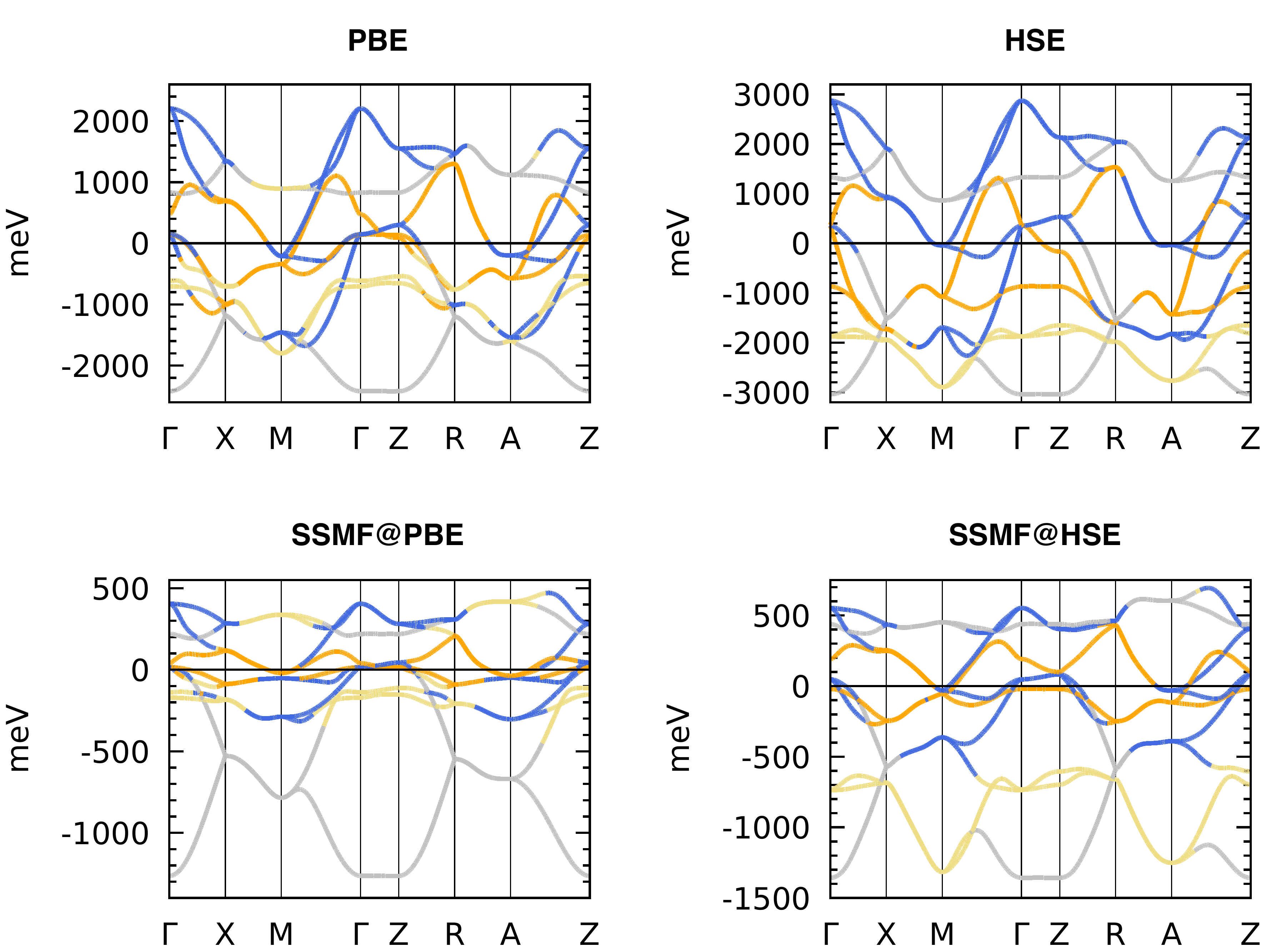}
\caption{Band structure of tetragonal FeSe in the four different approximations examined in this work. The colour corresponds to the major orbital character: $xy$ (orange), $xz/yz$ (blue), $z^2$ (gold) or $x^2-y^2$ (silver). The local interaction values are set to $U=4.2$\,eV, $J/U=0.20$ in the SSMF@PBE case and $U=5.0$\,eV, $J/U=0.20$ in the SSMF@HSE case.
\label{fig:bands-full}}
\end{figure}

%
% SSMF
%
The  quasiparticle renormalizations $Z_m$ and  on-site-energy  shifts $\tilde{\lambda}_m$ are  determined solving the self-consistent SSMF equations, which have been discussed in detail elsewhere~\cite{deMedici:2017}.
In this work, we computed the $\tilde{\lambda}_m =\lambda_m-\lambda^0_m$ parameters with the prescription used in Ref.~\cite{Yu:2012}. For a broader discussion, we refer the reader to Appendix~A of Ref.~\cite{Chatzieleftheriou:2020}.

%
% Local interaction
%
The local interaction values for the PBE reference Hamiltonian have been derived from \emph{ab initio} estimates by constrained Random-Phase Approximation (cRPA) of Ref.~\cite{Miyake:2010}, which for a purely d-model of bulk FeSe are $U=4.2$\,eV and $J/U=0.125$. 
It was however shown in detailed comparisons that the method we use in this work, slave-spin mean-field theory (SSMF), yields results slightly off from the more accurate dynamical mean-field theory (DMFT), and it was found as a general trend that the features obtained within SSMF with $J/U=0.20$--$0.25$ and the density-density Hund's Hamiltonian reproduce quite well those obtained within DMFT with $J/U=0.12$--$0.15$  and the Kanamori Hamiltonian, that includes spin-flip and pair-hopping terms (see Ref.~\cite{deMedici:2017}, Sec.~4.6.8).
The local interaction for the PBE reference Hamiltonian is therefore set to $U^{\rm PBE}=4.2$\,eV and $J^{\rm PBE}=0.20U^{\rm PBE}$, as done in~\cite{Villar-Arribi:2018}.

In the HSE case, we lack of cRPA estimates of the local interactions. 
%
%However, the HSE functional can be linked continuously to the PBE functional (both by reducing the screening length or the exact exchange fraction $\alpha$), whose correlation part is the same used in HSE as well. Since the $J/U$ is known to be less sensitive to screening~\cite{Werner:2016}, and the spread of the Wannier functions spanning the Fe-3d manifold increases only of $8\%$ with respect to the PBE case, we chose to keep the same $J/U$ value as in the PBE case.
%
Considering the systematic shift of the $J/U$ value due to the slave-spin approximation, we initially chose to keep the same ratio as in the PBE case of $J^{\rm HSE}=0.20U^{\rm HSE}$.
The on-site Coulomb repulsion has been then fixed to $U^{\rm HSE}=5.0$\,eV the value which roughly yields the same orbital differentiation between the $e_g$ and $t_{2g}$ as the SSMF@PBE case.
This choice is consistent with the general trend of Fock-like exchange resulting in overestimation of the bandwidth in similar compounds, as documented by GW+DMFT calculations~\cite{vanRoekeghem:2014,Hirayama:2017}, and with our main assumption of residing in the Hund's metal phase, as in can be seen by the quasiparticle renormalization value as a function of $U$ reported in the middle panels of Fig.~\ref{fig:FS-GZM}.
As a further check, in Fig.~\ref{fig:bands-UJ} we report the band structure of FeSe in both the PBE and HSE for different values of the effective on-site Coulomb repulsion $U$, showing that the topology of the $xy$ band is robust against its precise value and therefore a genuine feature of the non-local part of the self-energy.
The latter point can seen as well from the Fermi surface shown in Fig.~\ref{fig:FS-GZM}, computed along the plane passing through the $\Gamma$,$Z$ and $M$ symmetry points, where the outer $xy$ (orange) hole-pocket disappears only when screened Fock exchange is accounted for.

%
% Full band structure and z2
%
Finally, in Fig.~\ref{fig:bands-full} we report  the orbitally-resolved band structure of FeSe over the whole d-manifold bandwidth for the four different treatments of electronic correlations considered in this work. While the overall $t_{2g}$ description in the SSMF@HSE framework seems rather accurate, the $z^2$ bands (gold) lie  around $600-700$\,meV below $\varepsilon_{\rm F}$, in contrast to the $200-300$\,meV reported by ARPES~\cite{Reiss:2017}. 
Such a misplacement can be due to our choice of avoiding any fine tuning of the HSE screening via the $\alpha$ and $k_c$ parameters. Indeed from the inspection of the onsite energies $\epsilon_{im\sigma}$ the $z^2$ orbital results the most sensitive to the action of Fock exchange.

%===============================
%
\section{Fermi surface compensation}
%
%===============================

In the tetragonal phase of FeSe, three hole-like bands centered at the $\Gamma$ point (called $\alpha$, $\beta$ and $\gamma$) and two electron-like bands centered at $M$ point (called $\delta$ and $\epsilon$) are detected by ARPES experiments, all showing quasi two-dimensional character. 
At the zone centre, the top hole-like band ($\alpha$ band) forms a circular Fermi pocket~\cite{Shimojima:2014,Zhang:2015,Suzuki:2015,Fanfarillo:2016}, whereas the middle hole-like band ($\beta$ band) has its maximum in close proximity to the Fermi level, but there is no consensus whether it forms a hole pocket or not while dispersing along $k_z$~\cite{Watson:2015,Fanfarillo:2016}. A main $xz/yz$ character is attributed both to the $\alpha$ and $\beta$ bands by light-polarization analysis of the ARPES spectra. The lower hole-like band ($\gamma$ band) tops about $50$\,meV below the Fermi level and produces a fainter ARPES signal with respect to the other hole-like bands, which is explained by its main $xy$ character~\cite{Shimojima:2014,Nakayama:2014,Zhang:2015,Suzuki:2015,Fanfarillo:2016}.
The electron bands are both reported to form Fermi pockets at the zone corner, with the inner pocket of main $xz/yz$ character ($\epsilon$ band) and a fainter outer pocket of main $xy$ character ($\delta$ band)~\cite{Watson:2015,Fanfarillo:2016}. 

In Tab.~\ref{tab:fs-pockets} we report the Fermi-pocket sizes as measured along the $\Gamma$-$M$ and $Z$-$A$ directions in different ARPES experiments. The two $xz/yz$ pockets, the hole-like $\alpha$ and electron-like $\epsilon$, tend roughly to compensate one another in size in all the reported experiments. Considering that the $\beta$ hole-pocket contributes with a negligible number of carriers, the analysis of ARPES experimental data seem to indicate a non-compensated Fermi surface.
Two possible explanations might be an incoherent hole band crossing the Fermi level, which would be hardly detected by ARPES, or surface electron doping. The former scenario is supported by the prediction of a $xy$ hole pocket by single-site DMFT@PBE~\cite{Aichhorn:2010,Leonov:2015,Watson:2017}, which displays incoherence features in some regions of the parameter space~\cite{Aichhorn:2010}. This, however, would be at odds with the detection of the flat $xy$ band about 50~meV below $\Gamma$, which is sharply defined at low temperatures and does not display any particular temperature dependence upon heating, apart from a gradual loss of intensity~\cite{Shimojima:2014,Nakayama:2014,Zhang:2015,Suzuki:2015,Fanfarillo:2016}. 
Furthermore, the Fermi-liquid picture, once complemented with the screened Fock exchange as done in this work, seems capable of explaining the low-energy behaviour of tetragonal FeSe and the position of the $xy$ hole band at the zone centre, making us lean towards the electron doping scenario.
\begin{table}
\def\arraystretch{1.6}
\setlength{\tabcolsep}{12pt}
\centering
\begin{tabular}{c|ccccc}
                                                             &     $\alpha(\Gamma/Z)$   &   $\beta(\Gamma/Z)$     & $\gamma(\Gamma/Z)$   &   $\delta(M/A)$    &  $\epsilon(M/A)$  \\                      
\hline
Watson et al.~\cite{Watson:2015}        &     --/0.126                         &   --/0.052                       &  --                                    &  --                         &             0.057/--   \\
Fanfarillo et al.~\cite{Fanfarillo:2016}  &     0.06/0.12                       &   --                                 &  --                                    &  0.2/--                   &             0.07/0.1   \\
Yi et al.~\cite{Yi:2019}                         &     --                                    &   --                                 &  --                                    &  0.19/--                  &             0.1/--       \\
Huh et al.~\cite{Huh:2020}                   &     0.06/0.12                       &   --                                 &  --                                    &  0.17/--                 &             0.07/--      \\
\end{tabular}
\caption{Fermi pocket radii $k_{\rm F}$ as measured by ARPES along the $\Gamma$-$M$ ($Z$-$A$) direction in the tetragonal phase of FeSe. All the values are in $\AA^{-1}$.
\label{tab:fs-pockets}
}
\end{table}

%===============================
%
\section{Local spin-orbit effects}
\begin{figure}
\includegraphics[width=0.8\textwidth]{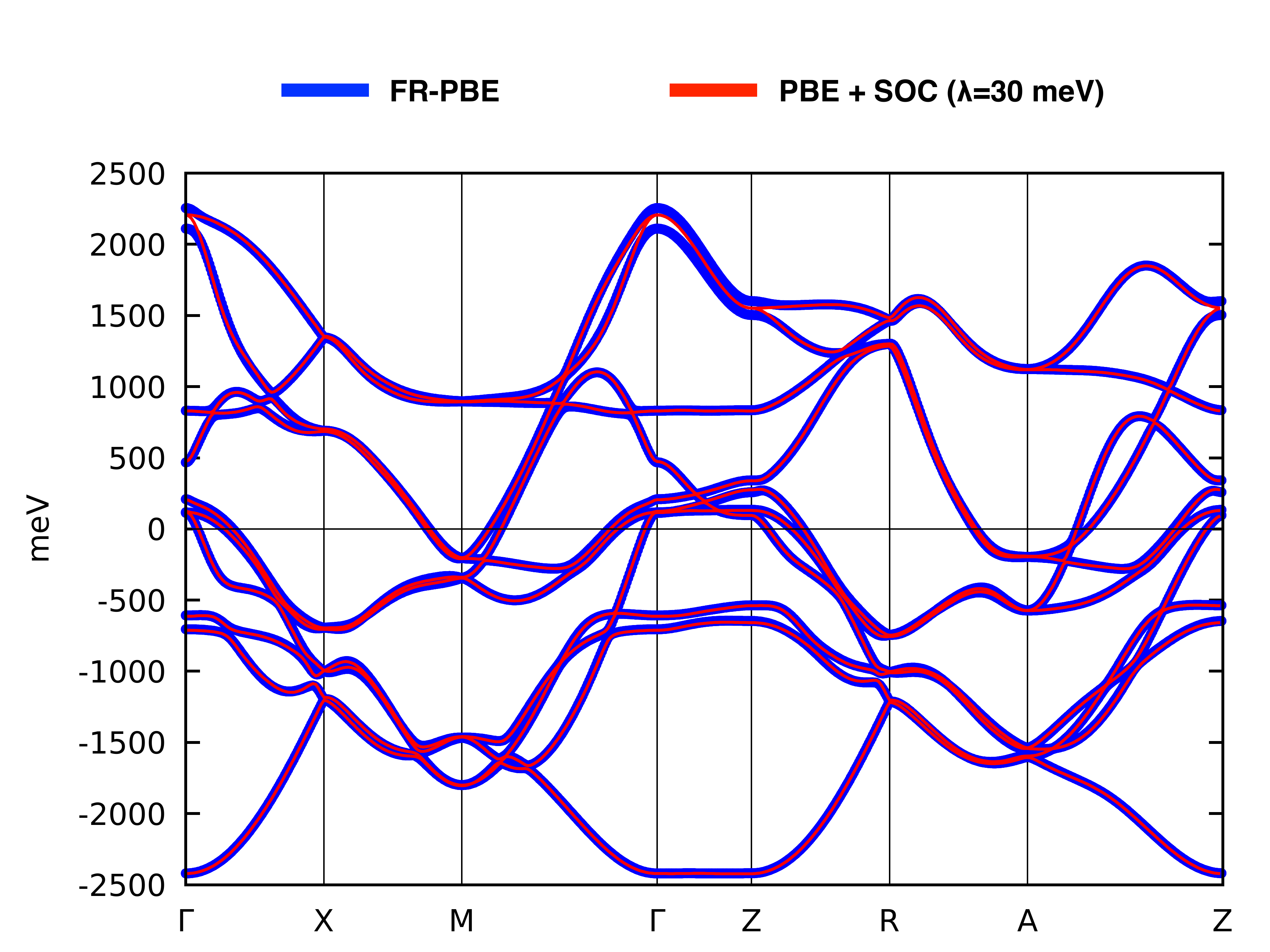}
\caption{%
PBE band structure of tetragonal FeSe computed using fully-relativistic pseudopotentials (FR-PBE, in blue) and adding local spin-orbit no top of the d-model obtained with maximally-localized Wannier functions (red).
\label{fig:SOC}
}
\end{figure}

%
%===============================
%
Local spin-orbit effects are included in the tight-binding formulation as
\begin{equation}
H_{\rm SOC}
=
\lambda \sum_{i} \hat{\mathbf l}_i \cdot \hat{\mathbf s}_i
\, ,
\label{eq:SOC}
\end{equation}
where $\hat{\mathbf l}_i$ and $\hat{\mathbf s}_i$ are the orbital and spin angular momentum operators acting at the $i$-th lattice site, respectively.
Wannier orbitals are assumed to be the eigenstates of $\hat{\mathbf l}_i$ with $l=2$, even though some $l=1$-character is present due to the hybridization between the $d$ and the $p$ manifold in FeSe.
By comparing with fully-relativistic DFT calculations, we find that an local spin-orbit term with strength $\lambda = 30$\,meV can reproduce the low-energy bands within the PBE approximation, as shown in Fig.~\ref{fig:SOC}. 
We notice that some higher-energy features, such as the splitting along the $\Gamma$--$Z$ direction, are not captured, most likely being due intersite spin-orbit effects.

When strong-correlations are taken into account, the local SOC term as the the one in Eq.~\eqref{eq:SOC} is added to the renormalized tight-binding obtained with SSMF. The Fermi level is recomputed at each given value of $\lambda$ in order to preserve the desired filling.

\bibliography{FeSe_hybrid.bib}